\documentstyle[twoside,fleqn,espcrc2,epsf]{article}

\title{
Truncated Overlap Fermions
}

\author{
A. Bori\c{c}i
\address{
Paul Scherrer Institute, CH-5232 Villigen PSI, Switzerland}
}
       
\begin{document}

\begin{abstract}
In this talk
I propose a new computational scheme with overlap fermions
and a fast algorithm to invert the corresponding Dirac operator.
\end{abstract}

\maketitle

\section{INTRODUCTION}

After many years of research in lattice QCD, it was possible
to formulate QCD with chiral fermions on the lattice
\cite{Kaplan,Shamir&Furman_Shamir,Narayanan_Neuberger,Hasenfratz_Laliena_Niedermayer}.

The basic idea is an expanded flavor space which may be seen
as an extra dimension with 
left and right handed fermions defined in the two opposite
boundaries or walls.

Let $N$ be the size of the extra dimension, $D_W$ the Wilson-Dirac
operator, and $m$ the bare fermion mass. Then, the theory with
{\it Domain Wall Fermions} is defined by the action
\cite{Kaplan,Shamir&Furman_Shamir}: 
\begin{equation}
\begin{array}{l}
S_{DW} := \bar{\Psi} {\cal{M}} \Psi = \sum_{i=1}^{N}
  \bar{\psi}_i [(D^{||}-1)\psi_i \\
  + P_{+}\psi_{i+1} + P_{-}\psi_{i-1}], \\
  P_{+}(\psi_{N+1} + m \psi_1) = 0, ~~P_{-}(\psi_0 + m \psi_N) = 0
\end{array}
\end{equation}
where $\cal{M}$ is the five-dimensional fermion matrix of the
regularized theory and
$D^{||} = M - D_W$ with $M \in (0,2)$ being a mass parameter.

In this talk I define
a theory with {\it Truncated Overlap Fermions} in complete
analogy with the Domain Wall Fermions by substituting
\begin{equation}
\begin{array}{l}
P_{+}\psi_{i+1} \rightarrow (D^{||} + 1)P_{+}\psi_{i+1} \\
P_{-}\psi_{i-1} \rightarrow (D^{||} + 1)P_{-}\psi_{i-1}
\end{array}
\end{equation}
while the boundary conditions remain the same as before.

Both theories can be compactified in the walls of the extra
fifth dimension as low energy effective theories (see below)
with the chiral Dirac operator $D$ satisfying the
Ginsparg-Wilson relation \cite{Ginsparg_Wilson}:
\begin{equation}\label{GWR}
\gamma_5 D^{-1} + D^{-1} \gamma_5 = 2a \gamma_5 R,
\end{equation}
where $a$ is the lattice spacing and $R$ is a local operator
trivial in the Dirac space (see below for $R$-locality tests).
From now on I set $a = 1$.

I defined Truncated Overlap Fermions such that in the large $N$
limit one obtains Overlap Fermions \cite{Narayanan_Neuberger}
with the Dirac operator given by \cite{Neuberger1}:
\begin{equation}
D_{OV} = \frac{1 + m}{2} - \frac{1 - m}{2} \gamma_5 \mbox{sgn}(H)
\end{equation}
where $H = \gamma_5 D^{||}$.

Until now computations with chiral fermions and
standard algorithms have been very
expensive. The extra fermion flavors introduce a large overhead.
One multiplication with the fermion matrix costs ${\cal{O}}(n)$
$D_W$-multiplications with $n \sim N$ for Domain Wall Fermions and
much larger for the overlap operator
\cite{Borici,Neuberger2,Edwards_Heller_Narayanan}.

In this talk I propose a fast algorithm
which makes these simulations an order of magnitude faster.
The key observation is the lack of gauge connections along
the fifth dimension.

\section{TRUNCATED OVERLAP FERMIONS}

I recall the action of the Truncated Overlap Fermions:
\begin{equation}
\begin{array}{l}
S_{TOV} := \bar{\psi}_1 [(D^{||}-1)\psi_1 \\
+ (D^{||}+1)P_{+}\psi_2-m(D^{||}+1)P_{-}\psi_N] \\
+ \sum_{i=2}^{N-1} \bar{\psi}_i [(D^{||}-1)\psi_i \\
+ (D^{||}+1)P_{+}\psi_{i+1}+(D^{||}+1)P_{-}\psi_{i-1}] \\
+ \bar{\psi}_N [(D^{||}-1)\psi_N \\
- m(D^{||}+1)P_{+}\psi_1+(D^{||}+1)P_{-}\psi_{N-1}]
\end{array}
\end{equation}

Let $P^T$ be the matrix representing the unitary transformation:
\begin{equation}
\begin{array}{l}
\chi_1 = P_{+} \psi_1 + P_{-} \psi_N, \\
\chi_i = P_{+} \psi_i + P_{-} \psi_{i-1}, ~~i = 2, \ldots, N
\end{array}
\end{equation}
and $S$ the matrix representing the diagonal transformation:
$\bar{\chi}_i = \bar{\psi}_i \gamma_5 (H - 1),  ~~i = 1, \ldots, N$.
Let also the transfer matrix along the fifth dimension be defined by:
$T = \frac{1 + H}{1 - H}$.

In the new basis I obtain the following action:
\begin{equation}
\begin{array}{l}
S_{TOV} =
\bar{\chi}_1 [(P_{+} - m P_{-}) \chi_1 - T \chi_2] \\
+\sum_{i=1}^{N-1} \bar{\chi}_i (\chi_i - T \chi_{i+1}) \\
+\bar{\chi}_N [\chi_N - T (P_{-} - m P_{+}) \chi_1]
\end{array}
\end{equation}
Integrating over the Grassmann fields I get:
\begin{equation}
\mbox{det} {\cal{M}} =
\mbox{det} [(P_{+} - m P_{-}) - T^N (P_{-} - m P_{+})]
\end{equation}
where I ignore the Jacobian factor coming from the diagonal
transformation.

If ${\cal{M}}_1$ is the same matrix
as  $\cal{M}$ but with the special choice $m = 1$,
I define the effective low energy theory with the
Dirac operator given by the equations:
\begin{equation}
D = (P^{T} {\cal{M}}_1^{-1} {\cal{M}} P)_{1,1},
~D^{-1} = (P^{T} {\cal{M}}^{-1} {\cal{M}}_1 P)_{1,1}
\end{equation}
where the subscript $1,1$ stands
for the $(1,1)$ block of an $N$x$N$ partitioned
matrix along the fifth dimension.

In terms of the transfer matrix the Dirac operator can be written as:
\begin{equation}
D = \frac{1 + m}{2} + \frac{1 - m}{2} \gamma_5 \frac{1 - T^N}{1 + T^N}
\end{equation}

I can repeat this derivation
for the Domain Wall Fermions with the sole changes
$\bar{\chi}_i = \bar{\psi}_i \gamma_5 (H P_{+} - 1),  ~~i = 1, \ldots, N$
and $T = \frac{1}{1 - H P_{+}}(1 + H P_{-})$,
the rest of the formulae remaining the same.

\begin{figure}
\vspace{.5cm}
\epsfxsize=6cm
\centerline{\epsffile[100 200 500 450]{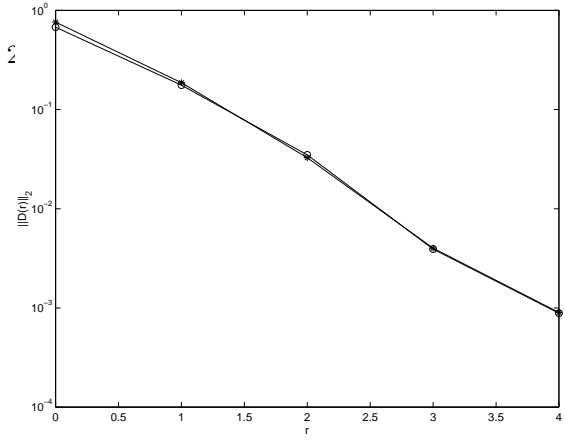}}
\vspace{-1cm}
\caption{
Norm of $D$ kernel in spin and color space with the distance $r$ from
the origin for $N = 4$ (circles) and $N = 32$ (crosses).
}
\vspace{-0.5cm}
\end{figure}
\begin{figure}
\vspace{.5cm}
\epsfxsize=6cm
\centerline{\epsffile[100 200 500 450]{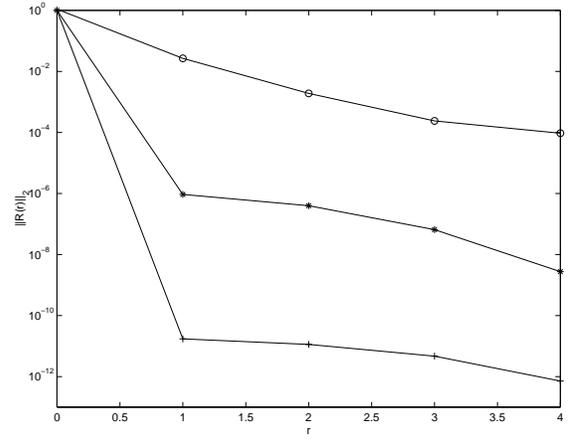}}
\vspace{-1cm}
\caption{
Norm of $R$ kernel in spin and color space with the distance $r$ from
the origin for $N = 4$ (circles), $N = 32$ (crosses) and $N = 64$ (pluses).
}
\vspace{-0.5cm}
\end{figure}

\section{LOCALITY AND GINSPARG- WILSON RELATION TESTS}

I test the Ginsparg-Wilson relation for the
Truncated Overlap Fermions, so that they can be used like
Domain Wall Fermions.

I computed the norm of $D$ and $R$ kernels in spin and color
space with the distance $r = \sqrt{x_{\mu} x_{\mu}}$ from the
origin on $4^4$ lattices at $\beta = 6.0$.
In Figs. 1 and 2 I show the maximum values at on-axis distances.
Note that $x_{\mu}, \mu = 1,2,3,4$ is measured modulo lattice size
in that direction.

Fig. 1 suggests exponential fall-off of $||D(r)||_2$, whereas Fig. 2
shows that $R$ approaches a Kronecker-Delta function as $N$ grows.

\section{A FAST INVERSION ALGORITHM}

I use Truncated Overlap fermions to define the following

{\it ALGORITHM1 (Generic) for solving the system $D_{OV} x = b$:}
\begin{equation}
\begin{array}{l}
\mbox{Given} ~N, ~x_0, ~r_0(=b), ~tol, ~tol_1, ~\mbox{set} ~tol_0 = 1 \\
\mbox{and iterate:} \\
for ~i = 1, \ldots \\
~~~~tol_0 = tol_0 tol_1 \\
~~~~{\bf \mbox{Solve} ~D y = r_{i-1} ~\mbox{within} ~tol_0} \\
~~~~x_i = x_{i-1} + y \\
~~~~{\bf r_i = b - D_{OV} x_i} \\
~~~~if ||r_i||_2 < tol, ~end ~for \\
\end{array}
\end{equation}
where by $o$ is denoted a vector with zero entries and $tol_1,tol$
are tolerances. $tol_1$ is typically orders
of magnitude larger than $tol$ such that the work per $D_{OV}$
inversion is minimized.

{\it Remark 1}. Bold face equations represent the smaller system
soultion and the correction of the right-hand side.
The straightforward application of the $ALGORITHM1$
gives a two-level algorithm. By calling it
again in solving the smaller system and iterating,
one gets a multi-level algorithm.

In Fig. 3 I compare the norm of the residual $r_i = b - D_{OV}x_i$
of the Conjugate Residual (CR) algorithm (which is optimal since $D_{OV}$
is normal \cite{Borici_thesis}) and $ALGORITHM1$. I gain about an order
of magnitude (in average) on $30$ $4^4$-configurations at $\beta = 6.0$
and $m = 0.1$.
For the coarse lattice I used $N = 6$ with the Truncated Overlap
Fermions and the Lanczos method to compute $D_{OV}$ \cite{Borici}.

{\it Remark 2}. Dynamical fermions can be implemented similarly.
The corresponding Hybrid Monte Carlo (HMC) algorithm can be
obtained by working with an approximate Hamiltonian in the coarse lattice
and by a global correction on the fine lattice.

One may use also as a starting point Truncated Overlap Fermions:
det$D = $det${\cal M}/$det${\cal M}_1$.
This way, all known simulation algorithms for dynamical fermions
apply.

\section{CONCLUSIONS}

I showed that Truncated Overlap Fermions may be used in two ways:

a) to implement Overlap Fermions in the same fashion as Domain
Wall Fermions;

b) to construct a multi-level inversion algorithm
for Overlap Fermions which saves an
order of magnitude of computer time compared to the state of the art
methods.

Further tests are needed to verify these results on larger lattices.

Recently, the possibility of a Multigrid algorithm along
all dimensions is raised \cite{Rebbi_LAT99}. In this case a
gauge fixing is needed.

\begin{figure}
\vspace{0.5cm}
\epsfxsize=6cm
\centerline{\epsffile[100 200 550 450]{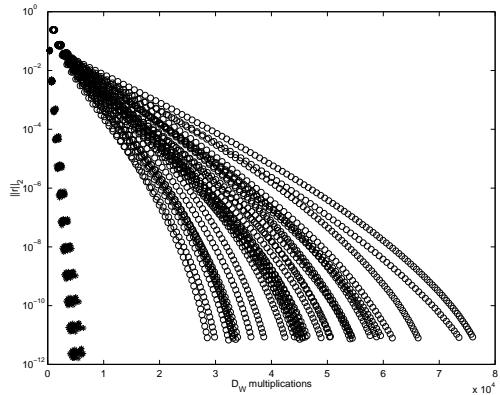}}
\vspace{-1cm}
\caption{Norm of the residual error vs. the number of $D_W$
multiplications on 30 configurations. Circles stand
for the straightforward inversion and stars for the new algorithm.}
\vspace{-0.5cm}
\end{figure}

\section{Acknowledgements}

The author would like to thank John Ellis for the hospitality at
CERN where these ideas initiated and for discussions on the
chiral fermions.

I would like to thank Philippe de Forcrand for suggestions
on how to improve the $ALGORITHM1$ which I will consider in the future.

The author thanks PSI where this work was done and SCSC Manno
for the allocation of computer time on the NEC SX4.

\end{document}